\documentclass{article}
\pdfoutput=1
\usepackage{arxiv}
\usepackage{amsmath}
\usepackage[utf8]{inputenc} % allow utf-8 input
\usepackage[T1]{fontenc}    % use 8-bit T1 fonts
\usepackage{hyperref}     
\usepackage{url}            % simple URL typesetting
\usepackage{booktabs}       % professional-quality tables
\usepackage{amsfonts}       % blackboard math symbols
\usepackage{nicefrac}       % compact symbols for 1/2, etc.
\usepackage{microtype}      % microtypography
\usepackage{cleveref}       % smart cross-referencing
\usepackage{graphicx}
\usepackage[numbers]{natbib}

\usepackage{makecell}
 
\usepackage{xcolor}

\usepackage{multirow}
\usepackage{bigstrut}
\usepackage{pgfplots}
\usepackage{multirow}
\usepackage{bigstrut}
\usepackage{subfig}
\usepackage{circledsteps}

\usepackage{xcolor}

\title{End-to-end Adaptive Distributed Training on  PaddlePaddle}

\date{}

\author{ 
	Yulong Ao$^1$\thanks{Both authors contributed equally to this research.}
	\And
	Zhihua Wu$^1$\footnotemark[1]
	\And
	Dianhai Yu$^1$
	\And
	Weibao Gong$^1$ 
	\AND
	Zhiqing Kui$^1$
	\And
	Minxu Zhang$^1$
	\And
	Zilingfeng Ye$^2$
	\And
	Liang Shen$^1$
	\AND
	Yanjun Ma$^1$
	\And
	Tian Wu$^1$
	\And
	Haifeng Wang$^1$
	\And
	Wei Zeng$^2$
	\And
	Chao Yang$^2$
	\AND
    $^1$ Baidu Inc.\qquad$^2$ Peking University
}

\renewcommand{\headeright}{}
\renewcommand{\undertitle}{}
\hypersetup{
	colorlinks,
	linkcolor={red!50!black},
	citecolor={blue!50!black},
	urlcolor={blue!80!black}
}

\begin{document}
\maketitle

\begin{abstract}
	Distributed training has become a pervasive and effective approach for training a large neural 
	network (NN)  model with processing massive data. However, it is very challenging to satisfy 
	requirements from various NN models, diverse computing resources, and their dynamic changes during a 
	training job. In this study, we design our distributed training framework in a systematic end-to-end view to 
	provide the built-in adaptive ability for different scenarios, especially for industrial applications and  
	production environments, by fully considering resource allocation, model 
	partition, task placement, and distributed execution. Based on the unified distributed graph and the 
	unified cluster object, our adaptive framework is equipped with a global cost model and a global planner, 
	which can enable arbitrary parallelism, resource-aware placement, multi-mode execution, 
	fault-tolerant, and elastic distributed training. The experiments demonstrate that our framework can satisfy 
	various requirements from the diversity of applications and the heterogeneity of resources with highly 
	competitive performance. The ERNIE language model with 260 billion parameters is efficiently trained on 
	thousands of AI processors with 91.7\% weak scalability. The throughput of the model from the 
	recommender system by employing the heterogeneous pipeline  asynchronous execution can be increased 
	up to 2.1 times and 3.3 times that of the GPU-only and CPU-only training respectively. Moreover, the 
	fault-tolerant and elastic distributed training have been successfully applied to the online industrial 
	applications, which give a reduction of 34.49\% in the number of failed long-term training jobs and an 
	increase of  $33.91\%$ for the global scheduling efficiency in the production environment.
\end{abstract}


\section{Introduction}
\label{sec:intro}
Distributed training with multiple or even thousands of machines has already become a pervasive and 
effective approach to train large neural network (NN)  models with massive data, which can 
significantly improve their generalization ability and prediction accuracy
\cite{kolesnikovBigTransferBiT2020, dosovitskiyImageWorth16x162021, 
	fedusSwitchTransformersScaling2021, megatron, TuringNLG17billionparameterLanguage2020, 
	raffelExploringLimitsTransfer2020, brownLanguageModelsAre2020, 
	naumovDeepLearningRecommendation2019, zhaoDistributedHierarchicalGPU2020}. 
However, the distributed training also brings more challenges into the framework design and implementation. 
First, models from various applications may need disparate parallel strategies 
based on their own characteristics. For example,  a typical recommendation model usually consists of a large 
sparse distributed embedding lookup layer and several dense fully connected layers, where the former incurs 
high data access cost but the latter is computationally intensive. Therefore the parallel strategies for
the recommendation models should be specially treated compared to other models.
Second, the distributed training for one model may require a different parallel strategy 
when targeting on a new architecture cluster with different devices or topology. This situation becomes 
more common since the available 
computing resources for users vary widely and even an organization usually has different types or 
generations of training devices. 
Third, there exists a strong interdependence between the model and the 
underlying resources in the context of distributed training. Specifically, the design of an efficient parallel 
strategy for one model should consider the topology and capability of used  resources while the dynamic 
change of used resources during a training job may require a better new parallel strategy for the same model. 
There is no doubt that a single parallel strategy can not fit into all the mentioned situations and 
it is also very challenging to manually design and implement individual strategies for all of them. 
So a good distributed training framework should address the following problem and provide the 
built-in adaptive ability for different scenarios:  \emph{how to adjust the parallel strategy automatically to 
	achieve efficient distributed training for various NN models, diverse computing resources, and even 
	the dynamic change of the used resources during a training job?}

\begin{figure}[!ht]
	\centering
	\includegraphics[width=0.54\linewidth]{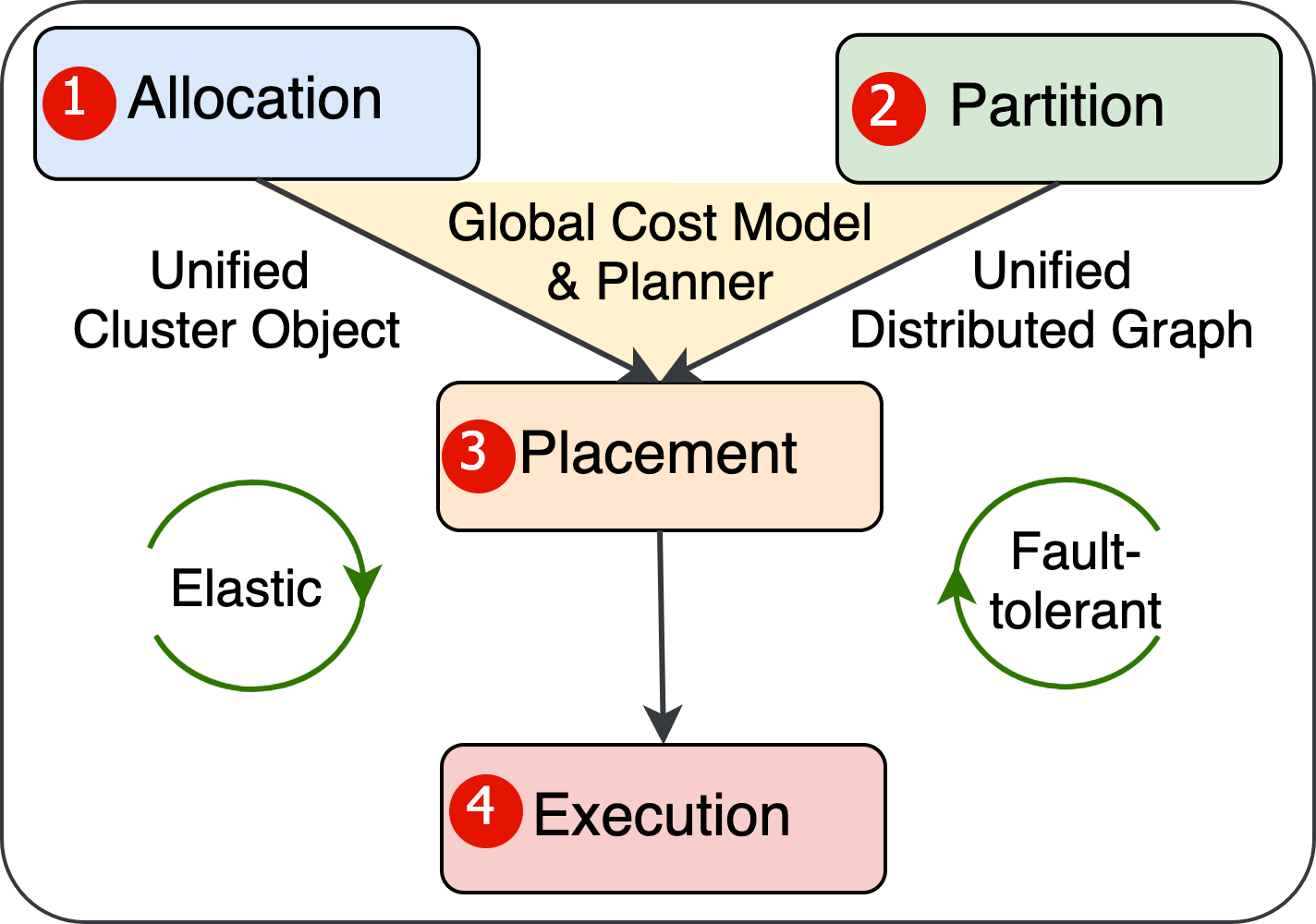}
	\caption{The essential steps of distributed training:  \Circled{1} allocate the  required resources,
		\Circled{2} partition the NN model,  \Circled{3} assign each partition to a specified device, \Circled{4}  
		execute tasks independently in a specific order with necessary communication.}
	\label{fig:design_principles}
\end{figure}

Unfortunately, most of the existing studies have limited adaptive ability. First, lots of parallel 
strategies have been proposed to partition the computation of NN models 
\cite{horovod, 
	liPyTorchDistributedExperiences2020a,rajbhandariZeROMemoryOptimizations2020,tensorflowmesh, 
	megatron,huangGPipeEfficientTraining2019, pipedream, 
	pipedream2bw,narayananEfficientLargeScaleLanguage2021a,
	liTeraPipeTokenLevelPipeline2021, hePipeTransformerAutomatedElastic2021}, but they are usually 
effective for specific models and need to be re-implemented manually for others. Although automatic 
parallelization  
\cite{flexflow,MindSpore,OneFlowWholeNew,gshard, xuGSPMDGeneralScalable2021} 
has been employed to minimize the efforts of developing parallelism for various NN models, most of them do 
not deal well with the heterogeneity of computing resources within a cluster or across clusters. 
Second, the device placement for the resulted partitioned tasks are often dealt with manually based on some 
heuristics from practice or the careful profiling results, so as the execution arrangement of the forward, 
backward 
and update computation. Even worse, there is almost no choice for users to select the appropriate distributed 
execution mechanism for their own workloads.  
Some studies \cite{pipedream, pipedream2bw, flexflow, pesto, deviceplacementrl, autosync, MindSpore, 
	caiTensorOptExploringTradeoffs2020, dapple}  try to build automatic planners or searchers to find the best
placement of execution policy, but they either only support very little parallelism or may fail to 
evaluate strategies more accurately with little consideration about the underlying hardware and the 
distributed execution runtime.  
Third, the existing frameworks usually delegate the resource management to external modules
\cite{tensorflow, pytorch, chenMXNetFlexibleEfficient2015, MindSpore, OneFlowWholeNew} 
or have insufficient awareness about it to make an adaptive and efficient distributed training when the 
resources change during a training job, especially for complicated parallelism
\cite{xiaoGandivaIntrospectiveCluster2018,gaoGAICentralizedTreeBased2018, 
	liSchedulingDistributedDeep2020, qiaoPolluxCoadaptiveCluster2021}.  Furthermore, very few of these 
studies cover all essential steps involved in the end-to-end distributed training process including 
resource allocation, model partition, task placement, and distributed execution, as shown in 
Figure~\ref{fig:design_principles}.

In this study, we design our distributed framework from a systematic end-to-end view to satisfy the 
mentioned versatile adaptive requirements by considering all the essential steps of the distributed training in 
Figure ~\ref{fig:design_principles}.  In the first place, a unified distributed graph is employed to represent 
arbitrary parallelism including all the existing parallelism and a unified cluster object is used to describe 
homogeneous 
and heterogeneous resources within a cluster or across clusters. Based on these two unified representations 
for parallelism and resources, a global cost model is developed to evaluate the cost of a distributed graph 
training on a specific cluster. Driven by the global cost model, we can utilize a global planner  to help us 
automatically choose 
better parallel strategies including partition and placement decisions according to the characteristics 
of the  given NN model and the cluster. To further give users more options for the distributed execution, a new 
asynchronous executor based on the actor model \cite{hewittUniversalModularACTOR1973} is added to our 
framework, in addition to the existing synchronous executor for collective communication and the 
asynchronous push-pull executor for parameter-server (PS) communication. 
Moreover, our framework has native support to 
realize the fault-tolerant and elastic distributed training in case of dynamic resource change for 
large-scale and long-time industrial scenarios.  Besides, we try to decouple these functions as many as 
possible and give advanced users  the maximum flexibility to select or configure the functions for their own 
sake. The main contributions are listed as follows:
\begin{itemize}
	\item An improved distributed graph based on the traditional computational graph is employed  by adopting
	three basic concepts: distributed tensor, distributed operator, and reshard transformation, which
	can represent arbitrary parallelism including all the existing parallelism.
	
	\item A unified cluster object is further built to represent the diverse resources for isolating the difference 
	between them within a cluster or across clusters. Based on the two unified representations, a global cost 
	model and a global planner are developed to select better resource-ware parallel strategies 
	automatically.
	
	\item A new distributed asynchronous executor based on the actor model is added to enrich our existing 
	distributed executors, which can automatically overlap the computation and communication as much as 
	possible and support different granularity and complex control flows. 
	
	\item The fault-tolerant and elastic distributed training are achieved through better built-in 
	interaction between  our distributed training framework and the platform scheduler, which can
	improve the overall resource utilization and make the fault-tolerant and elastic training of complicated 
	parallelism possible. 
\end{itemize}

Our adaptive distributed training framework is evaluated by various NN models on clusters of different 
architectures. The results show that our framework can train the popular GPT model with 146 billion 
parameters on 512 V100 GPUs and sustain $48.65\%$ of the theoretical peak FLOPS. The ERNIE 
language model  \cite{sun2021ernie}  with 260 billion parameters is efficiently trained on thousands of AI 
processors with 91.7\% weak scalability. In addition to these two NLP models, the large-scale image 
classification for face recognition can train 60 million classes on a  single node with 8 NVIDIA V100 GPUs and 
offer competitive or better performance than other frameworks. We also test the two classic 
models from the recommender system by employing the heterogeneous pipeline asynchronous execution 
and can obtain up to 2.1 times and 3.3 times the throughput of the GPU-only (only using GPUs of the servers ) 
and CPU-only  (only using the CPU servers)  training 
respectively, which can be further improved by automatic partition based on the cost model. Finally, the 
fault-tolerant and elastic distributed training can give a reduction of 34.49\% of 
failed long-term training jobs and an increase of $33.91\%$ for the global scheduling efficiency  in the 
production environment. 

\section{Design Principles}
\begin{figure}[!ht]
	\centering
	\includegraphics[width=0.6\linewidth]{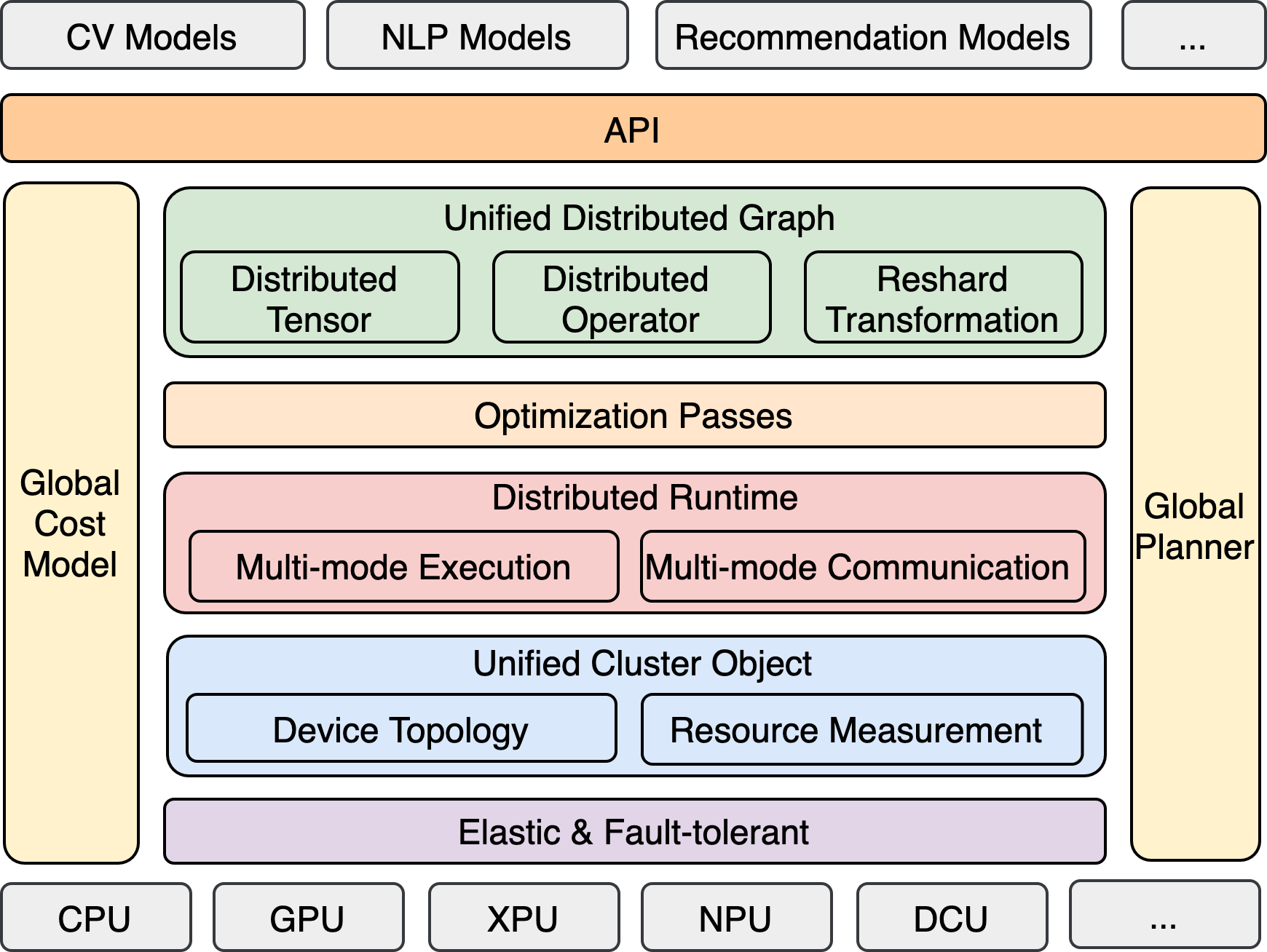}
	\caption{The architecture overview.}
	\label{fig:design_arch}
\end{figure}

In this study, three basic design principles are employed to handle the complexity of our framework to achieve
adaptive and efficient distributed training for requirements from different scenarios. First, 
the two unified representations for arbitrary parallelism and diverse resources are the core abstraction 
among all other implementations and optimizations. Next, different modules to implement the steps as 
mentioned in Section~\ref{sec:intro} are decoupled as many as possible to reach
maximum flexibility. Finally, the global end-to-end view for better overall performance is achieved by 
employing the global cost model and global planner. The architecture overview of 
our framework based on the design principles is shown in Figure~\ref{fig:design_arch}.

\label{sec:design}

\subsection{Unified Representation}
\label{subsec:rep}
As we have pointed out in Section \ref{sec:intro}, extensive research has been done to parallelize the 
computation of NN models. To deal with all the existing parallelism and other parallelism in the future, we build 
a general and 
unified distributed graph representation, which can describe arbitrary parallelism theoretically. The 
distributed graph can be seen as an enhancement to the traditional computational graph by adding more 
complete  parallelism semantics. One NN model can have different distributed graph representations, each of 
which encodes a unique parallel strategy but gives the same semantics as the serial computation. Ideally, the 
different distributed graphs only exert influence on the training performance without affecting the 
computation results.

For now, most of the existing studies mainly focus on clusters with homogeneous devices or 
heterogeneous devices within a cluster. However, the available computation resources from an organization 
are usually much more heterogeneous and may spread across different clusters with different types or 
generations of devices. In addition, some special-purpose accelerators for deep learning are 
employed to achieve significant performance improvements and power savings. These diverse heterogeneous
resources can be represented by an abstract and unified cluster object in our framework, which is globally 
visible to all involved processes. Different devices are modeled in the same way and can be distinguished by 
their types and generations. Additionally,  the cluster object can not only give the topology of all its connected 
devices but also abstract the ability of computation, storage, and communication in a unified 
quantitative approach.

\subsection{Maximum Flexibility}
\label{subsec:dec}
To have the versatile adaptive ability, the modules of the distributed training framework should be decoupled 
and can be configurable or changeable based on the different requirements. First, the distributed specification 
(i.e. the distributed graph) for a NN model is decoupled from the underlying implementation inspired by the 
work \cite{michaeledwardbauerLegionProgrammingDistributed2014,gshard, xuGSPMDGeneralScalable2021}. 
Users can configure how to partition the data and task since there is no universal partition 
policy for all NN models and they are familiar with the NN models they built. But it is the responsibility 
of the framework to provide necessary implementation mechanisms for performing the actual partition of NN 
models specified by users and automatically handle the necessary data movement and communication. So 
users can develop new partition strategies quickly without worrying about the tedious and 
error-prone implementation. And it is also very easy for us to improve and extend the distributed runtime 
mechanism while keeping the results consistent. 

In particular, the device-independent partition can be decoupled from the device-dependent placement. This 
can enable a specific partition to have a different placement policy not only on the same 
cluster but also a new one with different topology and devices. It can be 
observed that different placements for a partition, even on the same cluster, can exert a non-ignoble impact in 
terms of performance. For example, users can change the partition and placement for a better 
performance of the recommendation models based on their special model characteristics as mentioned in 
Section~\ref{sec:intro} and the given cluster. Unfortunately, only experts who know well about hardware 
architecture and parallel programming can make a good decision about the placement for a partition strategy. 
By decoupling these two aspects, most users can easily partition the NN model without considering the 
underlying hardware, while advanced users who care about optimal performance can replace 
the default placement policy with a tuned one.

\subsection{Global View}
\label{subsec:end}
To achieve the maximum flexibility needs to decouple the modules as many as possible, which is 
more likely to lose the global view and may lead to poor performance overall. However, it must be 
pointed out that the decoupling here is just a mechanism to realize the separation of concerned modules and 
should not hinder global optimization. Every module in our system is open to be configured or even
replaced as long as users conform to the interface between different modules. This 
means users have  full control over the system and can carry out application-aware or 
hardware-aware implementations or both of them to achieve a high-level performance depending on their 
knowledge of the application and the used resources. 

As shown in Figure~\ref{fig:design_arch}, our distributed framework consists of modules covering all 
of the essential steps mentioned in Section~\ref{sec:intro} to achieve an end-to-end global view.  While the 
most of existing studies focus on only parts of them, our work tries to provide a comprehensive solution by 
taking them all into consideration.  A global cost model is employed to estimate the required resources 
or evaluation of a specific placement to help us choose better policies. Moreover, driven by 
the cost model, a global planner is implemented to automatically find better policies. The unified 
distributed graph representation contains information about the partition and placement, and the abstract 
cluster object is also shared by the allocation and placement. Besides, the efficient elastic and fault-tolerant 
training is co-designed with other modules within our framework to satisfy the requirements for long-scale 
and  long-time distributed training in real industrial scenarios.

\section{Implementation}
This section details most of the modules of our distributed framework as shown in 
Figure~\ref{fig:design_arch}.  First, an improved distributed graph is constructed to represent 
arbitrary parallelism in a unified way. Then a cluster object is built to represent the diverse heterogeneous 
resources and a global cost model based on the two unified representations are developed to help us achieve 
resource-ware placement. After that, the multi-mode distributed execution including the new fully distributed 
asynchronous executor based the actor model is implemented to give users more options based on their own 
workloads. Finally, the fault-tolerant and elastic training is natively supported to adapt to the dynamic change 
of resources by better interaction between our distributed training framework and the platform scheduler.

\subsection{General Parallelization}
\label{subsec:para}

\begin{figure*}[ht]
	\centering
	\subfloat[The serial computational 
	graph]{\includegraphics[width=0.9\textwidth]{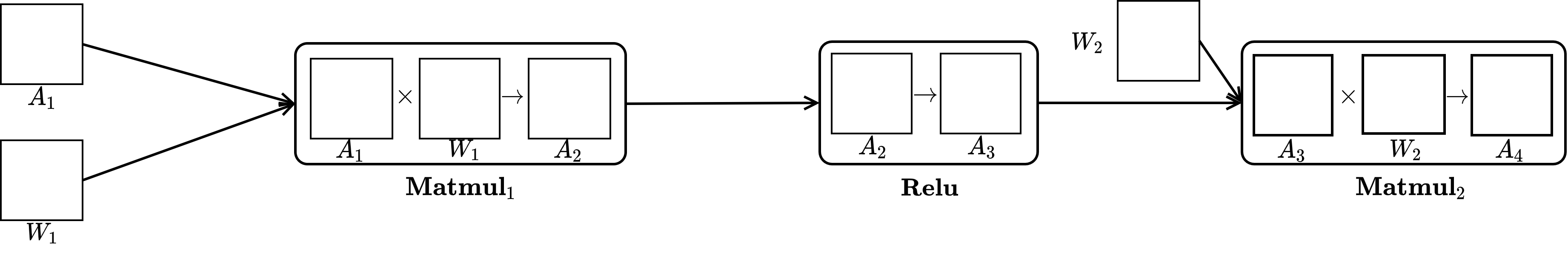}
		\label{fig:serial_comp_graph}}
	\hfil
	\subfloat[The distributed computational 
	graph]{\includegraphics[width=\textwidth]{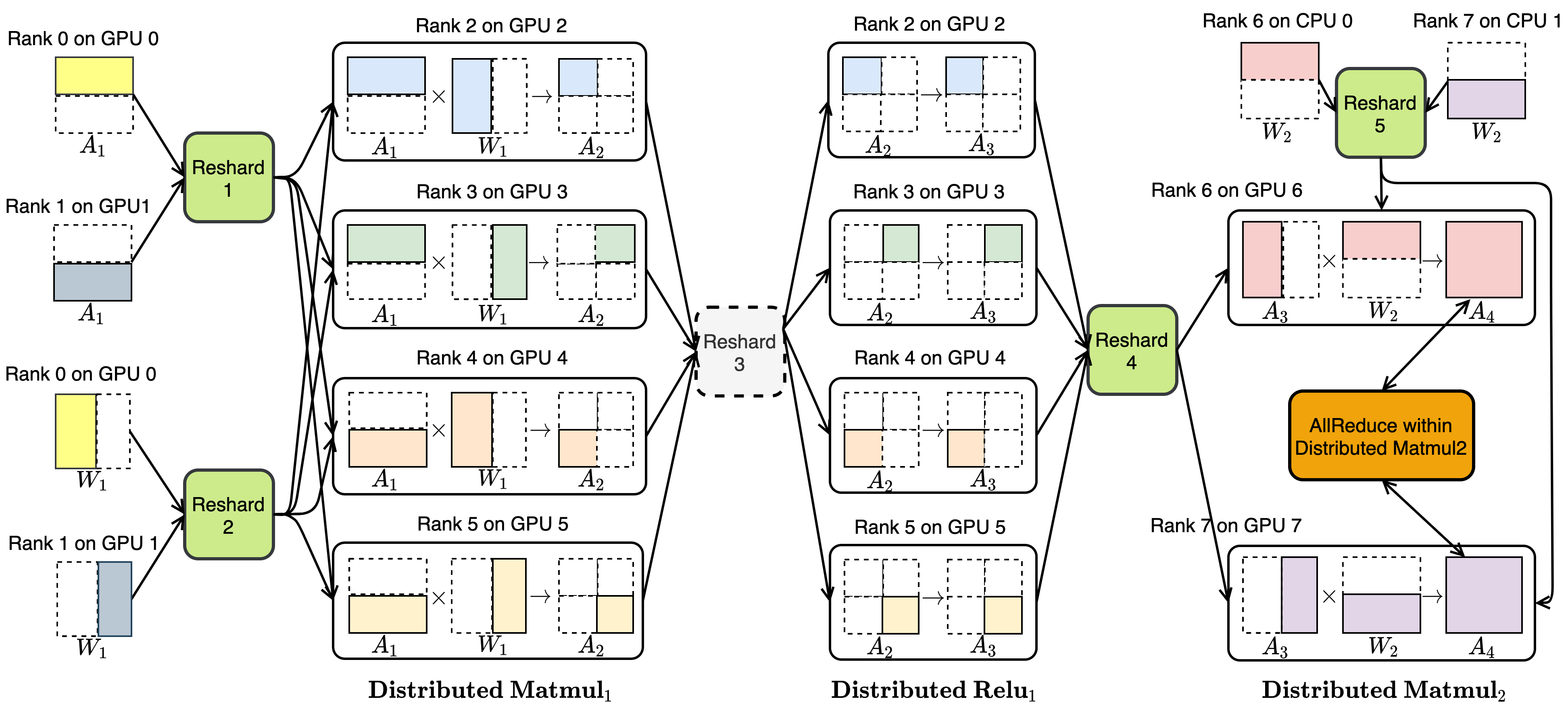}
		\label{fig:dist_comp_graph}}
	\caption{The transformation from the serial computational graph to the distributed computational graph 
		based on the distribute attributes in Table~\ref{tab:tensor_dist_attr} and Table~\ref{tab:op_dist_attr}. The 
		\texttt{Reshard 3} with a dashed gray box in this figure does nothing since the produced tensors and the 
		consumed requirements of the following operator are the same.}
	\label{fig:serial_dist_comp_graph}
\end{figure*}

To support all the existing parallelism and further explore potential ones, we reconsider the process of 
designing parallelism from a fine-grained perspective.  It is well known that the training process of a 
neural network can be described as a computational graph, where the computation at vertices are referred to 
as operators and the data that flows along edges is referred to as tensors. The basic idea is that the whole 
neural network's parallelism can be determined if the partition of each operator and tensor of it can be 
decided. To go a step further, all kinds of parallelism can be unified if the partition of operators and tensors 
can be described in a unified way. Fortunately, Some studies 
\cite{flexflow,MindSpore,OneFlowWholeNew,gshard, xuGSPMDGeneralScalable2021} have already adopted 
this fine-grained idea to achieve automatic parallelization. Based on these previous studies, we reformulate 
the traditional computational graph into a general distributed graph by introducing more parallelism 
semantics, which combines device-independent partition and device-dependent placement together. It is 
built on three basic concepts: distributed tensor, distributed operator, and reshard transformation. 
Figure~\ref{fig:serial_dist_comp_graph} illustrates a complete example of how to transform the serial
computational graph to our distributed graph by using these concepts.

\begin{table}[ht]
	\caption{The distributed attributes for $\mathbf{A_1}$, $\mathbf{W_1}$ and $\mathbf{W_2}$.   } 
	\label{tab:tensor_dist_attr}
	\centering
	\resizebox{0.6\linewidth}{!}{
		\begin{tabular}{@{}llll@{}}
			\toprule  
			\makecell[l]{\textbf{Distributed} \\ \textbf{Attributes}} & 
			\makecell[l]{$\mathbf{A_1}$ \\ (shape is [6, 8])}  & 
			\makecell[l]{$\mathbf{W_1}$ \\(shape is [8, 4])}  & 
			\makecell[l]{$\mathbf{W_2}$ \\(shape is [4, 4])} \\
			\midrule
			process\_mesh & [0, 1] & [0, 1] & [6, 7] \\
			\hline
			dims\_mapping & [0, -1] & [-1, 0] & [0, -1] \\
			\hline
			shard\_sizes & [[3, 3], [8, 8]] &  [[8, 8], [2, 2]] & [[2, 2], [4, 4]] \\
			\hline
			device\_placement & [GPU 0, GPU 1] & [GPU 0, GPU 1] & [CPU 0, CPU 1] \\
			\bottomrule
		\end{tabular}
	}
\end{table}

\begin{table*}[ht]
	\caption{The distributed attributes for $\mathbf{Matmul_1}$, $\mathbf{Relu_1}$ and $\mathbf{Matmul_2}$ 
	} 
	\label{tab:op_dist_attr}
	\centering
	\resizebox{\textwidth}{!}{
		\begin{tabular}{@{}llll@{}}
			\toprule  
			\makecell[l]{\textbf{Distributed} \\ \textbf{Attributes}} & 
			\makecell[l]{$\mathbf{Matmul_1}$}  & 
			\makecell[l]{$\mathbf{Relu}$}  & 
			\makecell[l]{$\mathbf{Matmul_2}$} \\
			\midrule
			Inputs & 
			\makecell[lt]{
				$\mathbf{A_1}$'s requirement: \\
				\quad process\_mesh: [[2, 3], [4, 5]] \\
				\quad dims\_mapping: [0, -1] \\
				\quad shard\_sizes: [[3, 3], [8, 8]] \\
				\quad device\_placement: [[GPU 2, GPU 3], [GPU 4, GPU 5]] \\
				$\mathbf{W_1}$'s requirement: \\
				\quad process\_mesh: [[2, 3], [4, 5]]\\
				\quad dims\_mapping: [-1, 1] \\
				\quad shard\_sizes: [[6, 6], [4, 4]] \\
				\quad device\_placement: [[GPU 2, GPU 3], [GPU 4, GPU 5]] \\
			} & 
			\makecell[lt]{
				$\mathbf{A_2}$'s requirement: \\
				\quad process\_mesh: [[2, 3], [4, 5]] \\
				\quad dims\_mapping: [0, 1] \\
				\quad shard\_sizes: [[3, 3], [4, 4]] \\
				\quad device\_placement: [[GPU 2, GPU 3], [GPU 4, GPU 5]] \\
			} & 
			\makecell[lt]{
				$\mathbf{A_3}$'s requirement: \\
				\quad process\_mesh: [6, 7] \\
				\quad dims\_mapping: [-1, 0] \\
				\quad shard\_sizes: [[6, 6], [2, 2]] \\
				\quad device\_placement: [GPU 6, GPU 7] \\
				$\mathbf{W_2}$'s requirement: \\
				\quad process\_mesh: [6, 7] \\
				\quad dims\_mapping: [0, -1] \\
				\quad shard\_sizes: [[2, 2], [4, 4]] \\
				\quad device\_placement: [GPU 6, GPU 7] \\
			}  \\
			\hline
			Ouputs & 
			\makecell[lt]{
				$\mathbf{A_2}$'s requirement: \\
				\quad process\_mesh: [[2, 3], [4, 5]] \\
				\quad dims\_mapping: [0, 1] \\
				\quad shard\_sizes: [[3, 3], [4, 4]] \\
				\quad device\_placement: [[GPU 2, GPU 3], [GPU 4, GPU 5]] \\		
			} & 
			\makecell[lt]{
				$\mathbf{A_3}$'s requirement: \\
				\quad process\_mesh: [[2, 3], [4, 5]] \\
				\quad dims\_mapping: [0, 1] \\
				\quad shard\_sizes: [[3, 3], [4, 4]] \\
				\quad device\_placement: [[GPU 2, GPU 3], [GPU 4, GPU 5]] \\
			} & 
			\makecell[lt]{
				$\mathbf{A_4}$'s requirement: \\
				\quad process\_mesh: [6, 7] \\
				\quad  dims\_mapping: [-1, -1] \\
				\quad  shard\_sizes: [[4, 4], [4, 4]] \\
				\quad  device\_placement: [GPU 6, GPU 7] \\
			} \\
			\bottomrule
		\end{tabular}
	}
\end{table*}

Each serial tensor in the original computational graph has one corresponding distributed tensor in the 
distributed graph, which is constructed based on its distributed attributes. It is worth emphasizing that 
the attributes here 
can represent an arbitrary partition of one tensor independent of the underlying hardware.  As shown 
in Table~\ref{tab:tensor_dist_attr}, the attributes give complete information about how to partition and place 
each shard of a tensor. The partition information is specified by the \texttt{process\_mesh}, 
\texttt{dims\_mapping},  \texttt{shard\_sizes}. The first two are very similar to the \texttt{device\_mesh} and 
\texttt{dims\_mapping} parameters in the \texttt{mesh\_split} interface of GSPMD 
\cite{xuGSPMDGeneralScalable2021}. However, we use the 
\texttt{process\_mesh} instead of the \texttt{device\_mesh} to decouple the logical partition from the physical 
placement and each process of the \texttt{process\_mesh} owns one shard of the original tensor. Another 
extra attribute \texttt{shard\_sizes} is introduced to specify the size of each shard along each dimension of the 
original tensor to enable more general parallelism such as the uneven parallelization. 
The last attribute \texttt{device\_placement} stores the physical placement 
information for each shard.  Like serial tensor, each serial operator also has a matched distributed operator. 
The behavior of a distributed operator is defined by the distributed attributes of its input and output tensors. 
The actual local computation of one distributed operator in each process uses the local operator and the local 
shards of its inputs and outputs belonging to that process. Although using the same representation, the 
distributed attributes for each input and output tensor within the operator only specify the requirement, 
which can be different from the distributed attributes of the actual tensor. This decoupling can support 
more advanced scenarios such as the operator being partitioned on some devices while the used tensors can 
be on others. Note that the \texttt{process\_mesh} 
must be shared by all its inputs and outputs in a distributed operator. The distributed 
tensor and operator across devices must preserve the semantics of their corresponding serial counterparts in 
a logical view. This enforcement will require necessary communications  (including data movement) to be 
inserted properly in the distributed graph, which can be classified into intra-operator communication and 
inter-operator communication.  The intra-communications are taken care of by the internal implementations 
of distributed operators to conform its serial semantic. The intra-communication is the responsibility of the 
reshard transformation to deal with the situation when a mismatch occurs between the required distributed 
tensor of a distributed operator and the actual distributed tensor. 

To put the above three concepts together, it can be concluded that the reformulated distributed graph 
can deal with arbitrary parallelism.  Based on it, we can also apply optimization passes to further improve the 
training performance. Moreover, the distributed graph representation makes us easy to provide different 
parallelization modes for users to satisfy their personalized requirements: some users can  annotate the 
distributed attributes of  selected tensors or operators and the rest ones will be automatically filled by our 
framework like \cite{gshard,xuGSPMDGeneralScalable2021}; others can use the global planner automatically 
to find the best distribute attributes for all tensors and operators.

\subsection{Resource-aware Placement}
\label{subsec:place}

\begin{figure*}[!th]
	\centering
	\includegraphics[width=0.98\linewidth]{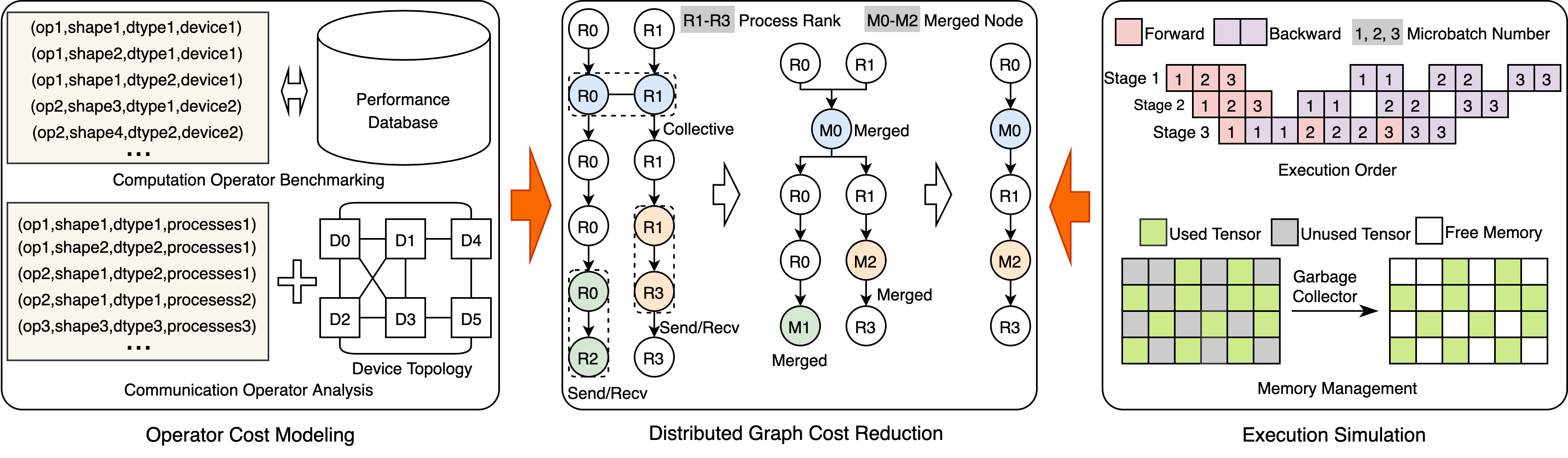}
	\caption{The distributed cost model based on the operator cost and the runtime simulation. Note that 
		each tuple of the lists in the left panel is the static information required for querying (or fitting) the 
		computation operator cost from the performance database or computing the communication operator 
		cost in an analytical way The information includes the operator type, the shapes of the inputs and 
		outputs, 
		the data types, the target device and the related processes, etc. }
	\label{fig:cost}
\end{figure*}

As described in Section~\ref{subsec:dec}, a partition of one NN model can have multiple placements 
onto the same cluster and each of them may reveal different performance. Our system can realize a 
resource-ware placement to achieve a high-level performance by employing two basic mechanisms. The 
first one is a cluster object to abstract the real underlying cluster for distributed training, which contains not 
only its topology information but also the quantitative measurements of its resources. The 
other mechanism we built  is an accurate global cost model to evaluate the cost of a distributed graph with a 
specific device placement based on the cluster object. Users can make a good decision to realize the 
resource-ware placement by utilizing these two mechanisms.

The challenge of constructing an abstract cluster object is that there are usually hundreds of machines 
connected through different topology networks from different clusters. This becomes more complicated 
when we want to train a NN model across clusters, especially with heterogeneous devices. Our cluster object 
is made up of some machine objects and each of these machines contains different components which can 
stand for any devices such as CPUs, GPUs, and link devices. This design allows a unified 
representation of the used clusters in different scenarios. Besides, we also store some quantitative 
measurements about the ability of computation, storage capacity, and communication inside a component 
object. For example,  the computation can be measured by FLOPS while 
the memories capacity is also stored.  The topology of the cluster is represented by a sparse adjacent matrix 
in a flattened way without distinguishing components in a machine or across machines. Each row or column of 
the matrix represents a component. The sparse matrix has an element to record the bandwidth and latency 
between two components if there is a connection between them.

In addition to the abstract cluster object, a global cost model is also developed, which can present the peak
memory consumption and execution time at the same time. The overall distributed graph cost is reduced 
based on the operator cost and the execution simulation as illustrated in Figure~\ref{fig:cost}. 
The computation operator cost is queried or fitted from the benchmarking performance database without  
adopting an analytical way since not only the shape of the inputs and outputs but also the kernel 
implementation and hardware architecture can affect the cost and they are often intertwined with each other 
intricately. 
However, to benchmark the  communication operator cost in a large scale, especially in complicated 
parallelism is impracticable because it needs to exclusively occupy all the resources for a long time. So the 
communication operators in our cost model are estimated analytically, which depend on not only the number 
of processes and message size but also the link topology and the utilization of these linkages from the built 
cluster object. After obtaining the cost of operators, the distributed graph cost can be inferred by applying the 
reduction rules to our distributed graph, which are similar to the studies 
\cite{jiaExploringHiddenDimensions2018,caiTensorOptExploringTradeoffs2020}.  Here we explicitly treat the 
communication as operators like computation including the collective and peer-to-peer ones while the 
previous studies usually use the edge cost to represent them. As a result, these communication operators can
be merged before applying the reduction. This merging process can easily retain the dependency among 
different ranks within the distributed graph and support more parallelism such as the intra-layer model 
parallelism. After merging communication operators, the key path will be identified by eliminating the 
branches and the final cost is the sum of all operators along that linear key path. Moreover, the execution 
order and memory management from the underlying framework runtime are also simulated in our cost model 
to make it more accurate since they have a strong impact on  the peak memory consumption and execution 
time.

\subsection{Multi-mode Execution}
\label{subsec:exe}
It is the responsibility of the executor in a deep learning framework to efficiently schedule and execute tasks 
on devices. To satisfy the different requirements,  our framework has already implemented the 
synchronous executor for collective communication and the asynchronous push-pull executor for 
PS communication. Recently, many studies
\cite{pipedream, pipedream2bw, narayananEfficientLargeScaleLanguage2021a} have shown that the 
scheduling order of forward and backward computation has a relatively large impact on both performance 
and memory utilization. In particular, it is extremely important  to overlap computation and communication as 
much as possible for distributed training to achieve high-level performance. Users have to manually arrange 
the scheduling order of the forward and 
backward computation to maximize the parallel degree and the 
overlapping of computation and communication. Fortunately, the actor model  
\cite{hewittUniversalModularACTOR1973}
proposed for distributed applications in the last century can automatically overlap the computation and 
communication as much as possible and inherently avoid race conditions. It has been first 
introduced by OneFlow \cite{OneFlowWholeNew} to implement the executor within its deep learning 
framework. To enrich our existing distributed executors, we also implement our own actor-like executor
which can support fine-grained task granularity and complex control flows.  With the help of our framework,
users can select an appropriate executor based on the workload of their NN models and even can combine
these executors together by exploring the characteristics of the different parts of one model.

The new actor-like executor also uses actors as the basic process units and each actor works in a physical 
thread and is responsible to execute the tasks assigned to it. And all actors are always residing in the runtime 
until the end 
of the training. Meanwhile, they are busy receiving messages from upstream actors, scheduling ready 
tasks to their corresponding devices, and sending messages to downstream actors.  It is often straightforward 
and easy to use one granularity to construct tasks such as each operator corresponding to a task. However, it 
is not suitable to use one granularity for all actors residing in different heterogeneous devices since the 
computation capabilities of these devices may vary widely. Our executor can bring users 
the flexibility to assemble one operator or multiple operators as a task based on the device's ability. It is also 
very convenient for us to deal with the complex control flow such as loop and condition since we can make 
the operators within these control flows into one big task.

In particular, we also applied heterogeneous pipeline asynchronous execution to the industrial recommender 
system, e.g.  training a large-scale CTR prediction, for using the heterogeneous computing resources 
efficiently. As mentioned in Section~\ref{sec:intro}, it is well known 
that a typical recommendation model 
usually contains one part which incurs high data access cost to convert massive high-dimensional sparse 
data into dense features, and the other part which is computationally intensive. So 
we can partition the model into different pipeline stages to enable efficient distributed training based on the 
characteristics of these layers, each of which may be a block of operators and serves as a task for the 
distributed execution.  These tasks are assigned to heterogeneous devices for better 
performance. Some of them are executed by the asynchronous push-pull executor while the others use the 
synchronized executor.  The partition strategy for the model and the used 
devices for each task can be configured by 
users as described in Section~\ref{subsec:rep}. The cost model mentioned in 
Section~\ref{subsec:place} can also be leveraged to guide the selection of  the partition, and the 
device allocation policy to further improve the overall throughput. 

\subsection{Fault-tolerant and Elastic Training}
\label{subsec:adaptive}

\begin{figure*}[!htb]
	\centering
	\subfloat[Original]{\includegraphics[width=0.35\textwidth]{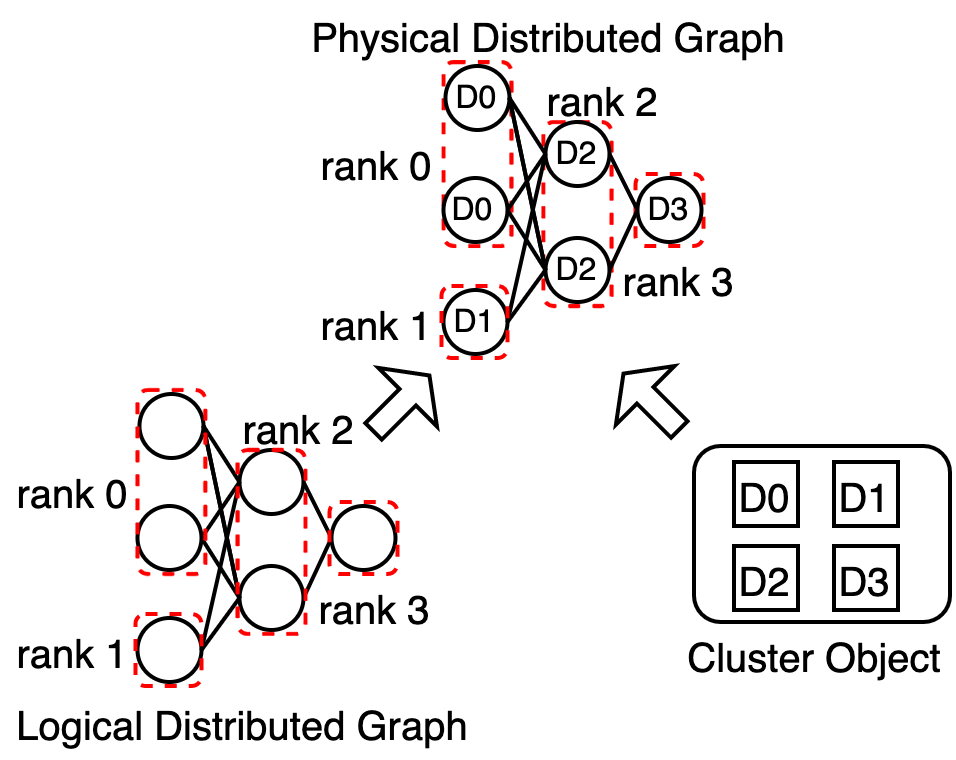}
		\label{fig:adaptive_original}}
	\hfil
	\subfloat[Fault-tolerant]{\includegraphics[width=0.35\textwidth]{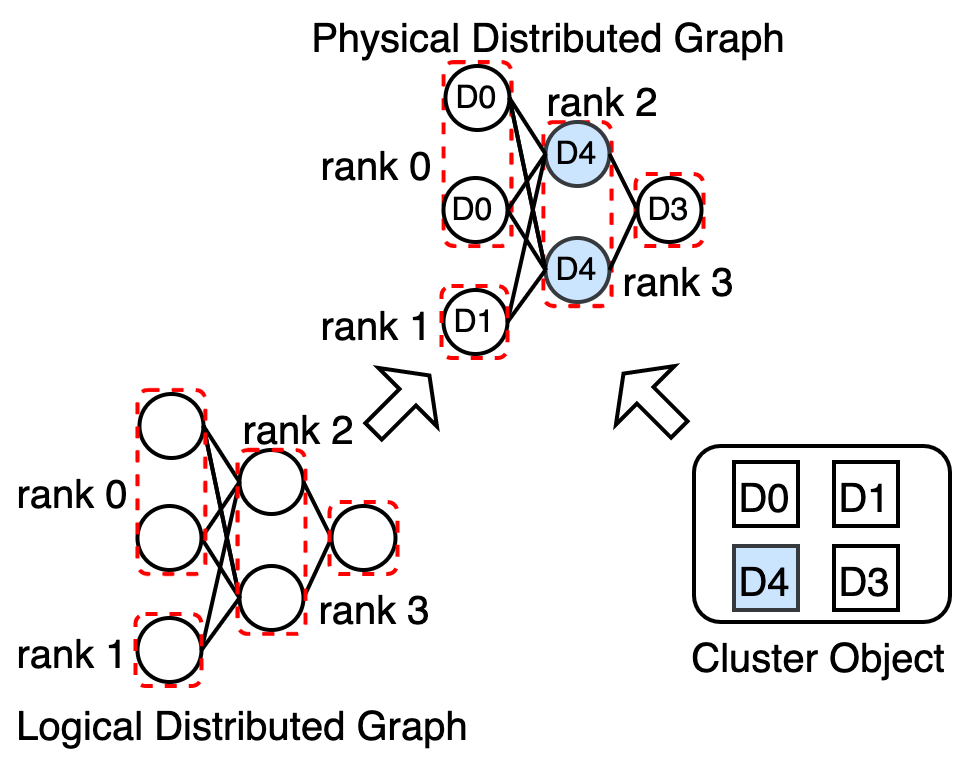}
		\label{fig:adaptive_fault}}
	\newline 
	\subfloat[Scale down]{\includegraphics[width=0.35\textwidth]{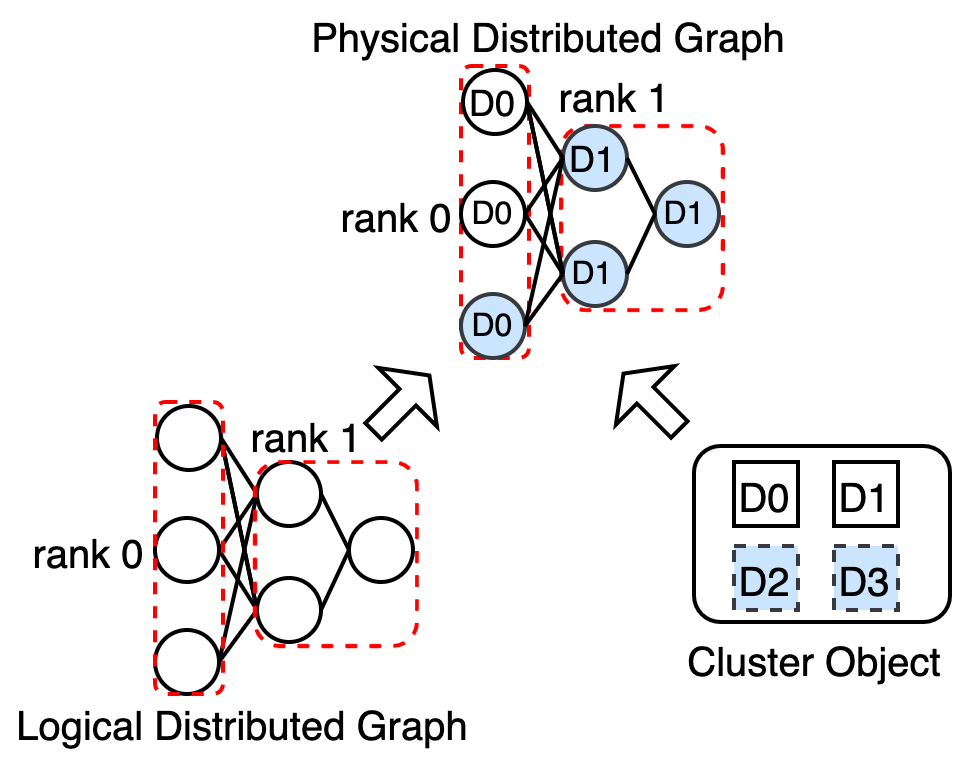}
		\label{fig:adaptive_scale_down}}
	\hfil
	\subfloat[Scale up]{\includegraphics[width=0.35\textwidth]{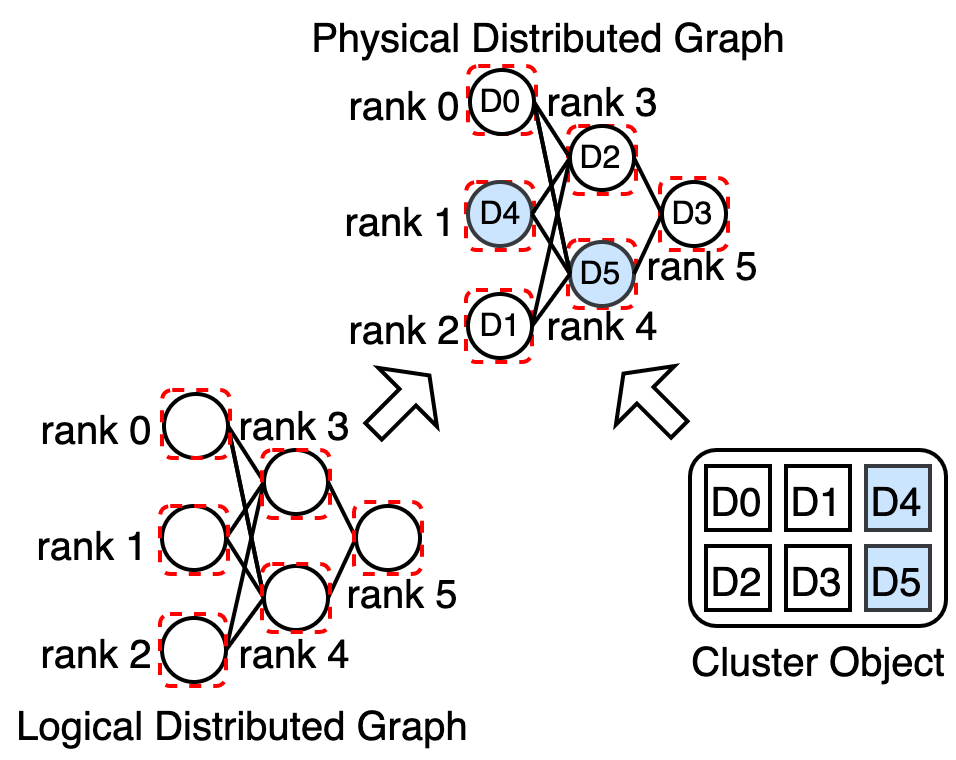}
		\label{fig:adaptive_scale_up}}
	\caption{Different cases supported by the fault-tolerant and elastic training: a) the original 
		training with four devices; b) the fault-tolerant training replaces the failed device D2 with D4 without
		re-partitioning the graph; c) the elastic training releases the devices D2 and D3 with re-partitioning the 
		graph; d) the elastic training with adding the devices D4 and D5 with re-partitioning the graph.}
	\label{fig:adaptive}
\end{figure*}

The limited awareness between the distributed training framework and the platform scheduler impedes the 
full exploration of the resource utilization in a large-scale and long-time training, especially on a 
multi-tenant industrial platform. To achieve efficient and robust adaptive training, more interactions 
will be needed between the distributed training framework and the platform. Based on this idea, we provide 
the built-in support for fault-tolerant and elastic training, which is illustrated by Figure~\ref{fig:adaptive}. The 
job scheduler from the platform usually tries its best to allocate resources in a more aggregated way when a 
resource request arrives. However, lots of fragments will also be produced if jobs with various occupancies 
are collocated in the same node over time. With the help of  the fault-tolerant and elastic training from our 
framework,  the job migration can be automatically and efficiently performed if the current available 
resources do not meet the desired one.  In addition, a preemption operation will be more efficient  for 
other jobs with a high priority of resources allocation, which makes some of the low-priority jobs 
stop early and these jobs will be resumed later for better resource utilization. 

The basic fault-tolerant training can be implemented by reloading the recently saved checkpoint when there 
occurs a failure. However, we should make a trade-off between benefit and cost about how to save and 
reload a checkpoint to achieve efficient fault tolerance. In general, a checkpoint  includes two parts, the 
weights of the model and the states of the training progress and hype- parameters, etc. Here we adopt a 
three-level checkpoint strategy to reduce the global overhead: 1) a save-before-exit hook triggered before 
shutting down when an exception is caught internally or a job is intended to restart manually; 2) a fast 
in-memory backup on each device for a fast partially recovery where the saving interval is calculated by the 
benefit-cost ratio; 3) a snapshot of weights and states in the persistent storage taken by all devices with a 
larger interval, which can provide both partial and full recover capability with a much lower cost. This strategy 
is also optimized in several ways. First of all, the states are saved at a high availability key-value backend for 
fast sharing among all devices. Second, the checkpoint process takes place asynchronously  for 
higher efficiency. Furthermore, each device uses an adaptive strategy based on the parallelization to 
reduce unnecessary saving and loading, especially in complicated parallelism. For example, the 
devices working as the data parallel role can save their weights in turn alternatively and the devices working 
as the model parallel role only save their own weights.

The elastic training can make better utilization of resources for a multi-tenant platform. However, existing 
distributed frameworks provide limited support to realize the elastic training for complicated
parallelism. Thanks to the decoupling of our unified parallelization and resource-aware placement, one 
serial graph can have different distributed graphs adapting to the available resources. For convenience, we 
define a distributed graph with specified resources as a scheme. Before training, the granularity of the 
change level about resources which will automatically trigger the elastic training needs to be configured. In 
the beginning, the distributed training job employs the initial scheme based on the available resources. A 
scaling up is performed if the next level of more resources for a candidate scheme is satisfied, while a scaling 
down is performed when the platform recycles resources of the current job to other jobs with high priority. 
In brief, the scaling in our elastic training between different schemes works in a discontinuous way gradually. 
Note that the hyperparameters such as learning rate and batch size may be coordinated to avoid causing 
harm to the model accuracy.  And the saved weights may also need to be automatically converted when 
changing from one scheme to another especially using different parallelism.

\section{Case Studies and Evaluation}
In this section, we try to give a complete evaluation of our adaptive distributed framework using experiments 
of the various NN models from different applications. First, the general parallelization is tested on GPT, ERNIE 
\cite{sun2021ernie} and the classification task for face recognition on large scale. Next, an MLP model 
and a multi-view DNN model for recommendation system by using heterogeneous pipeline asynchronous 
execution are compared with the CPU-only and the GPU-only training respectively. And these 
two models are further used to show the ability of our cost model. Last, the fault-tolerant and 
elastic training are evaluated for long-time distributed training in the production environment.

\subsection{Large-scale Parallelization}
\label{subsec:hybrid}

\begin{table*}[ht]
	\centering
	\caption{Results for GPT models on V100 GPUs in the different configurations.}
	\resizebox{\textwidth}{!}{
		\begin{tabular}{ccccccccccc}
			\toprule
			\makecell{\textbf{Parameters(B)}} & 
			\makecell{\textbf{Attention} \\ \textbf{heads}} & 
			\makecell{\textbf{Hidden} \\ \textbf{size}} & 
			\makecell{\textbf{Layers}}& 
			\makecell{\textbf{MP}} & 
			\makecell{\textbf{PP}} & 
			\makecell{\textbf{DP}} & 
			\makecell{\textbf{GPUs}} & 
			\makecell{\textbf{Batch size}} & 
			\makecell{\textbf{Achieved TFLOPS} \\  \textbf{per GPU}} & 
			\makecell{\textbf{Percentage of} \\ \textbf{theoretical peak FLOPS}} \\
			\midrule
			9.66B & 16 & 4096 & 
			48 & 8 & 1 & 
			1 & 8 & 512 & 
			55.77 & 44.61\% \\
			
			16.31B & 16 & 6144 & 
			36 & 8 & 2 & 
			2 & 32 & 512 & 
			60.84 & 48.67\% \\
			
			91B & 32 & 8192 & 
			112 & 8 & 16 & 
			1 & 128 & 512 & 
			62.05 & 49.64\% \\

			146B & 32 & 11264 & 
			96 & 32 & 8 & 
			2 & 512 & 2048 & 
			60.82 & 48.65\% \\
			\bottomrule
		\end{tabular}%
	}
	\label{tab:weakscaling}%
\end{table*}%

\textbf{GPT results on NVIDIA GPUs}.
We first train GPT, which is one of the most popular language models, on V100 GPUs (32 GB) by combining 
data parallelism (DP), intra-model parallelism (MP) and pipeline model parallelism (PP). The results of the 
throughput with different configurations are shown in Table \ref{tab:weakscaling}. 
From the table, it can be seen that our system can achieve a minimum of 44.61\% of the theoretical peak 
FLOPS 
in all configurations. And the model with 146 billion parameters  in the table can be trained by 512 
V100 GPUs and sustains $48.65\%$ of the theoretical peak FLOPS. 

\textbf{ERNIE results on NPUs from Peng Cheng Laboratory}.
In addition to GPT, another well-recognized language model ERNIE \cite{sun2021ernie} is trained on
NPUs on a large scale by employing the hybrid parallelism. Unlike GPT, the last transformer layers of 
ERNIE have smaller shapes, which results in a big performance difference between GPU and NPU. 
Fortunately, by conducting the resource-ware training, our framework takes on a parallel strategy different 
from GPU architecture, which can give $2.19$ times speedup from 392 NPUs to 480 NPUs and give $2.17$ 
times speedup from 1568 NPUs to 1920 NPUs as shown in Table  \ref{tab:npusmall}. 
And all parallel strategies can achieve good weak scalability of throughput when we 
increase the DP degree and fix the MP and PP degrees. Finally, ERNIE can be scaled up to 
260 billion parameters on 1920 NPUs with 91.7\% weak scalability by using the resource-ware 
configuration.

\begin{table}[ht]
	\centering
	\caption{Comparison between the two different parallel strategies when  training ERNIE on NPUs. The 
		one is the same as GPUs while the other one is resource-aware for NPUs.}
	\resizebox{0.6\linewidth}{!}{
		\begin{tabular}{cccccc}
			\toprule  
			\makecell{\textbf{Parallel strategies}} & 
			\makecell{\textbf{NPUs}}  & 
			\makecell{\textbf{DP}} &
			\makecell{\textbf{Global} \\ \textbf{batch size}} &
			\makecell{\textbf{Speedup}} \\
			\midrule
			\multirow{2}[1]{*}{\makecell{Same as GPUs  \\ (260 billion parameters)}} & 392    & 1     & 512   & - 
			\\
			& 1568    & 4     & 2048   & - \\
			\hline
			\multirow{2}[2]{*}{\makecell{Resource-aware \\(260 billion parameters)}} & 480    & 1     & 512   & 2.19 \\
			& 1920    & 4     & 2048   & 2.17 \\
			\bottomrule
		\end{tabular}%
	}
	\label{tab:npusmall}%
\end{table}%

\begin{figure}[ht]
	\centering
	\includegraphics[width=0.8\linewidth]{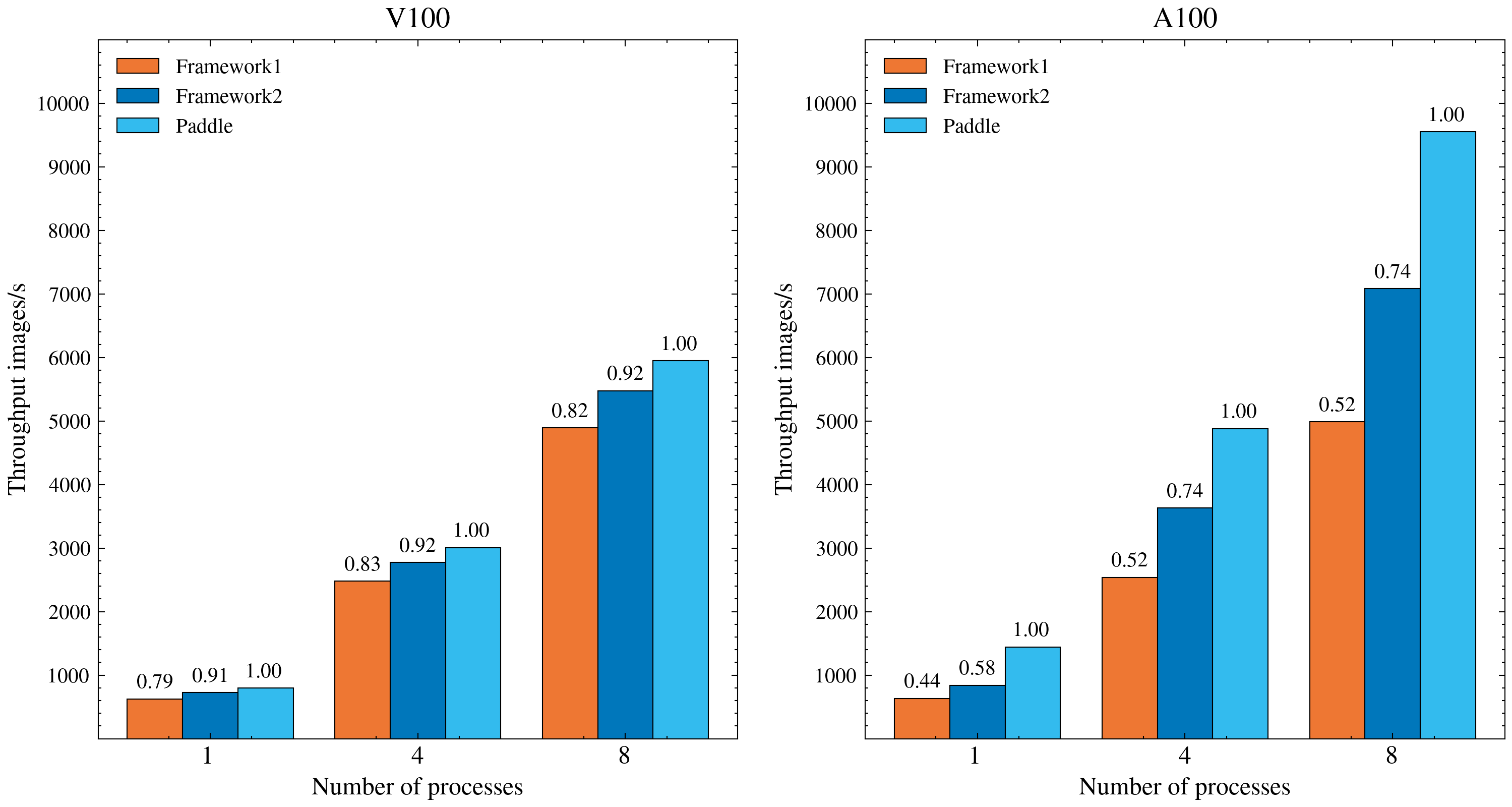}
	\caption{Large-scale Image Classification based on ResNet50.}
	\label{fig:plsc}
\end{figure}

\textbf{Image Classification results on NVIDIA GPUs}.
The open-source large-scale image classification tool for face recognition \cite{plsc} also uses the hybrid 
parallelism of our framework and can support 60 million classes on a  single node with 8 NVIDIA 
V100 GPUs. The results on V100 and A100 GPUs are shown in Figure~\ref{fig:plsc}, where the 
backbone is ResNet50. In these tests, the batch size is 128 per process, the sample ratio is 0.1,  and 
the number of classes is 93431. We can see that not only the results of 
FP32 and FP16 on V100 but also those on A100 deliver stronger performance than other 
frameworks. In particular, in the case of FP16 on A100, our framework can achieve about two 
times speedup compared to others at most.

\subsection{Heterogeneous Pipeline Asynchronous Execution}
\label{subsec:heterps}

\begin{figure}[ht]
	\centering
	\includegraphics[width=0.6\columnwidth]{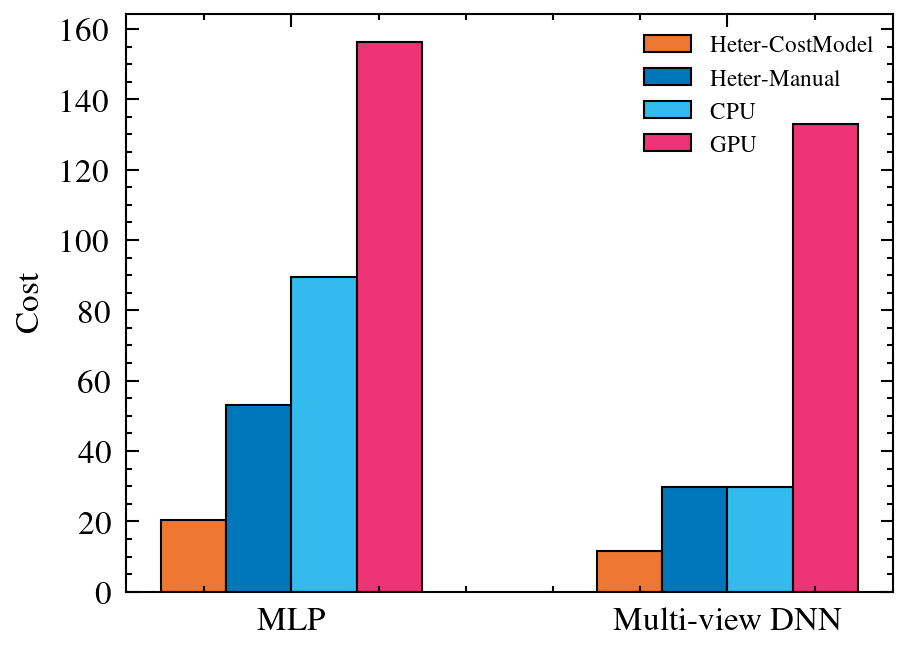}
	\caption{Cost comparison with different partition strategies.}
	\label{fig:partition_strategy}
\end{figure}

This experiment below is used to show the advantage of our heterogeneous pipeline asynchronous  
execution (Heter), especially for the recommendation models as described in Section~\ref{subsec:exe}. A 
MLP model and a 
multi-view DNN model for the click through rate (CTR) prediction problem 
\cite{zhangDeepLearningBased2019} are trained by using an open dataset Criteo. We compare the 
costs of the GPU-only training, CPU-only training, Heter training configured with the cost model, and Heter 
training configured manually, where the costs are used for training in terms of the resource prices from our 
cloud platform.
The manual Heter training partitions these models into two stages with an embedding layer in the 
first stage and the remaining parameters in the second stage while the partition strategy searched by the 
cost model further splits the last several fully connected layers.  As shown in 
Figure~\ref{fig:partition_strategy}, the 
partition strategies searched by the cost model for both models have a lower cost than the CPU-only, 
GPU-only, and manually configured Heter training.

\begin{figure}[ht]
	\centering
	\includegraphics[width=0.8\columnwidth]{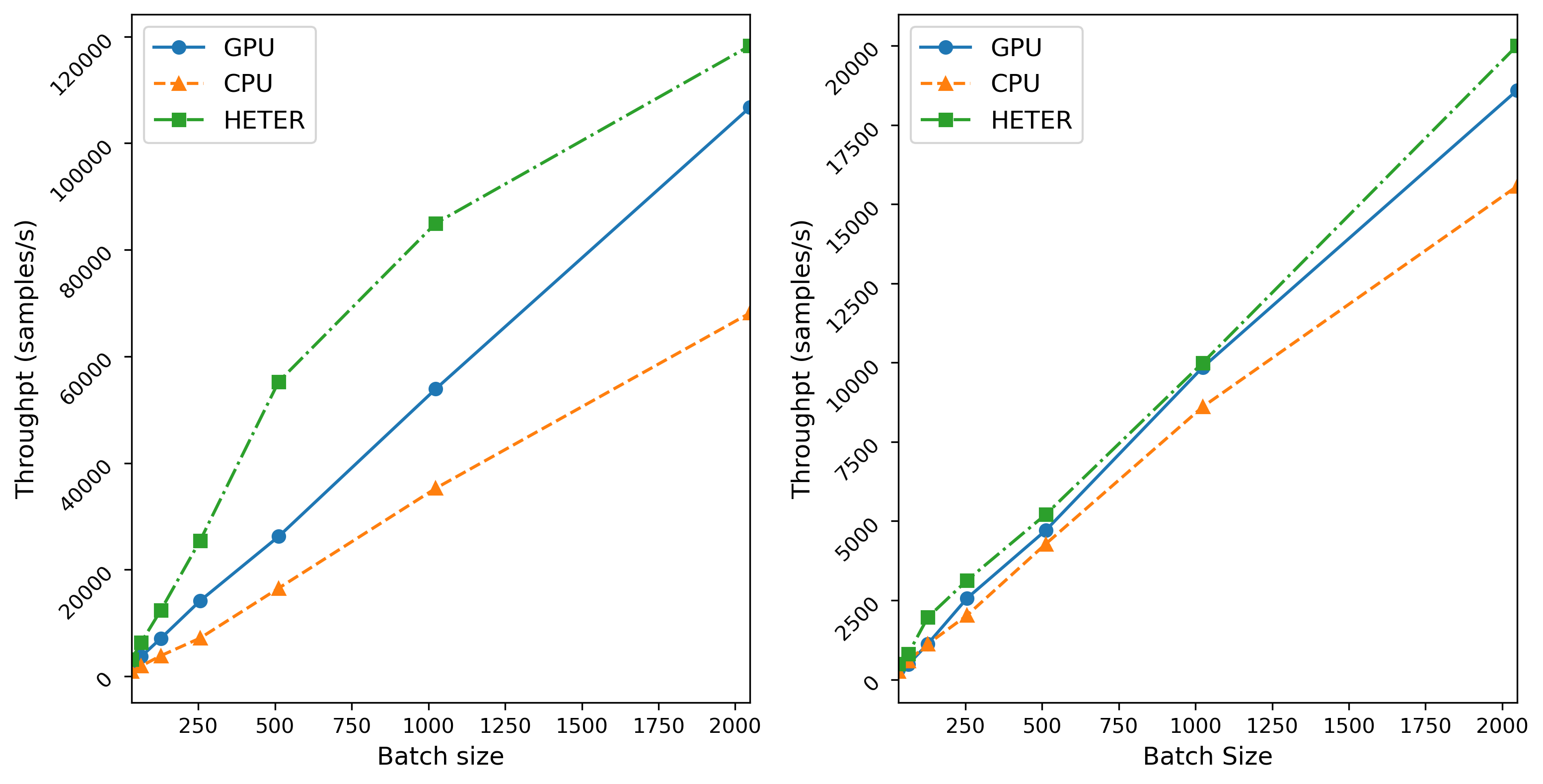}
	\caption{Throughput comparison with different batch sizes in the MLP (left) and multi-view DNN model 
		(right).}
	\label{fig:ctr_dnn_2emb}
\end{figure}

\begin{figure}[ht]
	\centering
	\includegraphics[width=0.6\columnwidth]{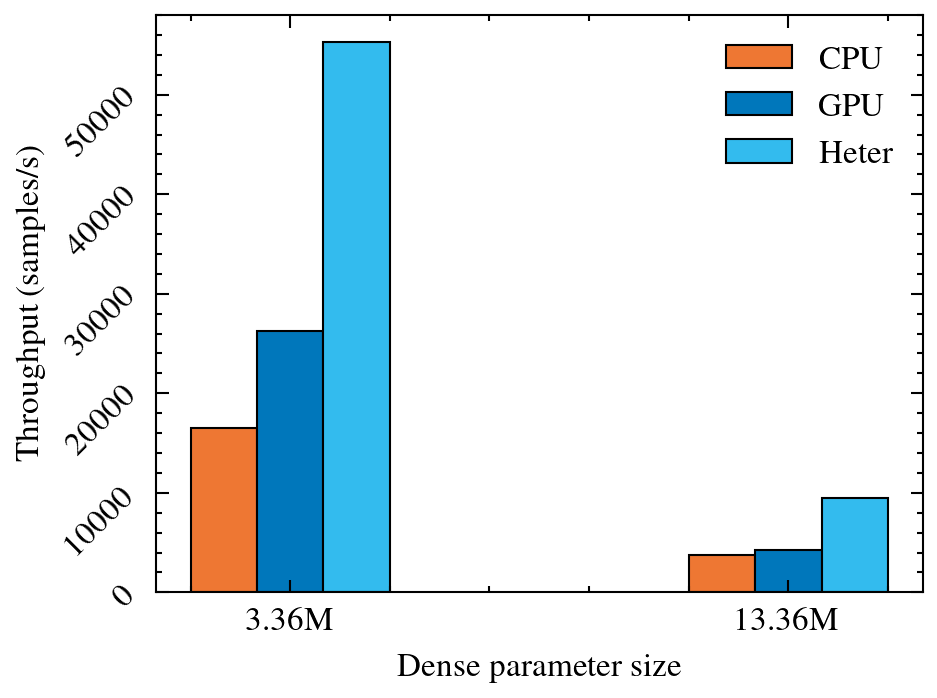}
	\caption{Throughput comparison with different dense parameter sizes in the MLP model.}
	\label{fig:dense_model_size}
\end{figure}

We also change the batch size and dense layers' size to prove the advantage of  the  Heter 
training. Here we try to compare it with
the GPU-only training and the CPU-only training where the former uses one GPU server with 8 NVIDIA V100 
GPUs and the latter uses 10 CPU servers, where the price of one GPU server is roughly equal to the price of 
10 CPU servers. The  Heter partition the models into two stages with an embedding layer in the 
first stage and dense layers in the second stage, where the former uses 5 CPU machines and the latter uses 
4 GPUs (the price of those resources are equal to these of the GPU-only training and CPU-only training). First, 
We fix dense layers' size and increase the batch size to compare the training throughputs of 
the three training methods. Figure \ref{fig:ctr_dnn_2emb} shows that all the throughputs of the three 
methods are growing with the increase of the batch size, and the Heter training
shows higher throughput than others in all batch sizes. When the batch size is 512, the 
Heter can achieve 2.1 times and 3.3 times the throughput of the GPU-only and 
the CPU-only training for the MLP model, while the Heter training can 
achieve 1.3 times and 1.4 times for the multi-view DNN model.  Then the batch size will be fixed to 512 and 
the  throughputs are compared by using different dense layers' sizes in the MLP model. As shown in Figure 
\ref{fig:dense_model_size},  the throughputs of all three types of training are decreasing when increasing the 
dense model size. However, the Heter has an apparent advantage because the 
communication can be well overlapped with the computation, which can give 2.2 times the throughput of the 
GPU-only training and 2.5 times the throughput of the CPU-only training at most. 

\subsection{Fault-tolerant and Elastic Training}
\begin{figure}[ht]
	\centering
	\includegraphics[width=0.6\columnwidth]{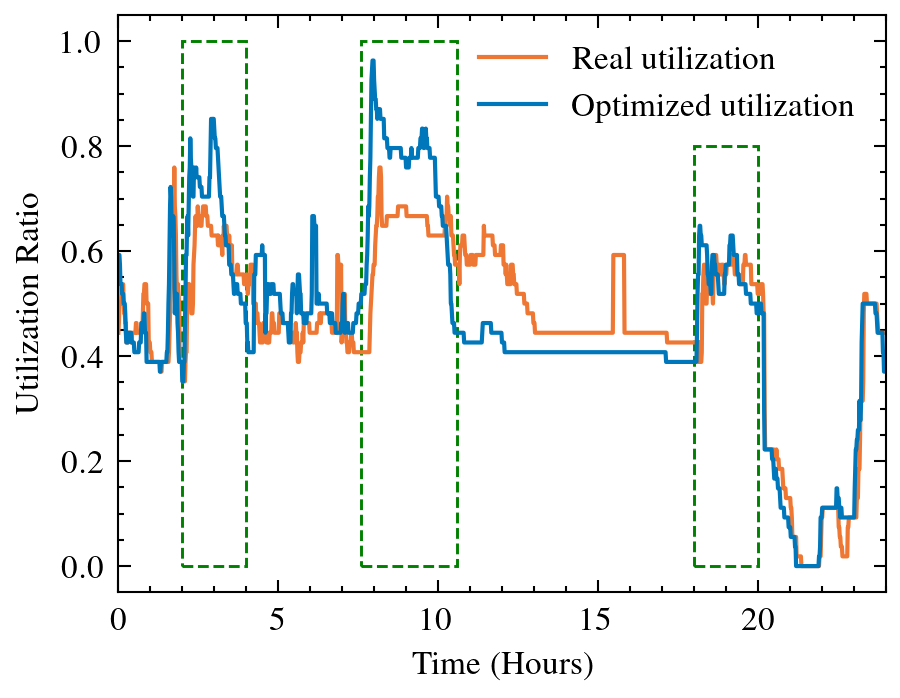}
	\caption{GPU utilization with the job migration}
	\label{fig:job-migration}
\end{figure}

Without fault tolerance, all the resources that a job holds before being terminated
will be wasted in case of failure. We estimate the fault-tolerant training of our framework under our production 
environment for one month. The results shows it can reduce $34.49\%$ of failed long-term 
training jobs to be rerun, which can further save scheduling and queuing time. It is very hard for traditional job 
schedulers to make an optimal scheduling decision since they have little
information about the training workload.  With the help of our system, the job migration can be also easily 
conducted when it is suitable as mentioned in Section~\ref{subsec:adaptive}. 
It can improve the utilization efficiency by $33.91\%$ by estimating jobs in the production environment for a 
day long.  In practice, most jobs are delayed for several minutes after being submitted
even without restricting the resources occupation, 
while they can be scheduled  immediately by performing the job 
migration (the line of optimized utilization), which can improve the efficiency of the global job scheduling.
More specifically, as shown in the rectangle zone of Figure \ref{fig:job-migration},
the utilization rate is improved by applying the job migration so that enough resources are available for an 
immediate scheduling.
In contrast, you may note that in some periods such as the center of the figure, the optimized utilization is 
lower than the real one,
it also makes sense because the original queued jobs are consumed earlier which can result in a higher global 
throughput.

\section{Related Work}
\textbf{Different Parallelism.}
Lots of efforts have been made to partition the computation of NN models to achieve efficient large-scale 
distributed training. Horovod  \cite{horovod} is known for its high performance in data parallelism with the 
optimized \texttt{AllReduce} implementation. The sharded data parallelism 
\cite{xuAutomaticCrossReplicaSharding2020} can shard 
the weight update computation across replicas. And this idea is extended by ZeRO 
\cite{rajbhandariZeROMemoryOptimizations2020} which shards both the 
weight parameters and other optimizer states. The following ZeRO-Infinity 
\cite{rajbhandariZeROInfinityBreakingGPU2021}
uses the offload technique and NVMe to train very large models on a small number of GPUs. Megatron 
\cite{megatron} introduces model 
parallelism to train large transformer-based language models by splitting some layers, 
and Mesh-TensorFlow \cite{tensorflowmesh}  can also support intra-layer model parallelism with an 
interface for easily 
specifying parallelization strategies. Pipeline model parallelism is another common
parallelization by splitting large models into different stages. GPipe \cite{huangGPipeEfficientTraining2019} 
partitions a batch into small 
microbatches to reduce the pipeline bubble size without changing the strict synchronous optimizer 
semantics. PipeDream \cite{pipedream} and its successor PipeDream-2BW \cite{pipedream2bw} relax
the synchronization by  
maintaining different versions of weights and updating stale weights in an asynchronous way to improve 
the pipeline efficiency. TeraPipe \cite{liTeraPipeTokenLevelPipeline2021} proposes a fine-grained pipeline 
parallelism across tokens for 
auto-regressive models. PipeTransformer \cite{hePipeTransformerAutomatedElastic2021}  can 
automatically 
adjust the pipeline and data parallelism by freezing some layers and allocating resources for the remaining 
active layers. Deepspeed \cite{DeepSpeedExtremescaleModel2020} and the work 
\cite{narayananEfficientLargeScaleLanguage2021a} 
combine data parallelism, intra-layer model parallelism, and inter-layer pipeline parallelism to train extremely 
large models and the latter is optimized to give better 
performance. The unified distributed graph adopted by our framework can represent all the mentioned
parallelism and can also support other parallelism in the future.

\textbf{Automatic Parallelization.} 
FlexFlow \cite{flexflow} proposes the SOAP representation to express partitioning schemes and can 
automatically search a fast scheme for a specific cluster by using an execution simulator to 
accurately predict the performance for a scheme.  PipeDream 
\cite{pipedream} and  PipeDream-2BW \cite{pipedream2bw} can automatically search the possible
pipeline parallelism configurations and the cost models of them are analytical methods based on profiling 
results. Dapple \cite{dapple}  further
improves PipeDream's planner by allocating different numbers of devices to each pipeline stage.
Pesto \cite{pesto} can fast find the optimal placement and scheduling by formulating the problem as 
an integer program for model parallelism.  Instead of adopting an analytical way, Mirhoseini et al. 
\cite{deviceplacementrl} employs the reinforcement learning method to automatically decide the placement 
for each operator of the computational graph.  Autosync \cite{autosync}  also learns to optimize 
synchronization strategies to lower the bar for data parallelism based on a model- and 
resource-dependent representation. With regard to operator cost,  TVM \cite{tvm} utilizes a 
learning-based cost model for evaluating operator implementations. Besides, the deep learning 
frameworks \cite{MindSpore,OneFlowWholeNew,gshard, xuGSPMDGeneralScalable2021} can also support
automatic parallelization for NN models to minimize efforts of developing complex parallelism. Our framework 
can also automatically parallelize NN models for arbitrary parallelism 
based on the unified distributed graph, and our search space is larger according to the distributed 
attributes, which combine the partition and placement information together. And the global cost 
model can synthesize the analytical and machine-learning methods as well as the runtime information at 
the same time.

\textbf{Fault tolerant and elastic training.} 
Recently, the failure tolerant and elastic functions for NN model training on large clusters and in cloud 
environment have been studied in some works.
Wu et al. \cite{wu2021elastic} propose a lightweight coordination layer between the cluster scheduler 
and the deep learning framework to enable elasticity with a simple API.  Ma et al. \cite{ma2021towards} 
adopt the mixed-integer programming model to maximize the training progress in real a production 
environment. Saxena et al. \cite{saxena2020effective} formulate the batch size selecting problem into a 
fast dynamic programming problem and solves it in real-time. Hu et al. \cite{hu2021optimal} propose an 
unbiased algorithm to deal with the heterogeneity in the cluster by changing the volatility of parameters in the 
model and utilizing incomplete local updates. KungFu \cite{mai2020kungfu}, a library proposed by Mai et 
al. introduces the ability to synchronize hyperparameters in order to make a distributed machine learning 
adaptive. Besides, several cluster schedulers are proposed in consideration of the NN workload. 
Gandiva \cite{xiaoGandivaIntrospectiveCluster2018} exploits the predictability of iterations in NN training to 
time-slice GPUs efficiently across multiple jobs, thereby delivering low-latency and  improving cluster 
efficiency. GAI \cite{gaoGAICentralizedTreeBased2018} is a centralized 
tree-based scheduler which can allow jobs to preempt resources occupied by other lower priority jobs.  
DeepSys \cite{ liSchedulingDistributedDeep2020} considers 
resource allocation and task placement to provide efficient job scheduling based on a speed 
model and memory model.
Pollux \cite{qiaoPolluxCoadaptiveCluster2021} can improve the scheduling performance by adaptively 
co-optimizing inter-dependent 
factors both at the per-job level and at the cluster-wide level. In this study, we try to provide native support 
in our framework for better interaction with the cluster scheduler to achieve adaptive training.

\section{Conclusion and Future Work}
This study shows that our adaptive distributed training framework designed in the global end-to-end 
view can satisfy various requirements from the diversity of applications and the heterogeneity of resources 
and give a competitive and high-level performance.  In the future, we will explore and support the adaptive 
requirement from the dynamic change of the NN model during a training job. The unified distributed graph will 
be improved to become a distributed intermediate representation, and the current manual optimization 
passes may be re-implemented by utilizing the existing compiler techniques. For now, the communication is 
implemented by selecting collective or peer-to-peer primitives from different backend libraries and can be 
also unified in the same interface, which can assure callers of the transparent use  despite the underlying 
cluster. Based on the unified communication, it is much easier for the 
distributed training  to explore the available resources from different organizations.  Besides, we will further 
enhance our framework by collaborating more with the resource platform to achieve intelligent 
scheduling and transparent resource allocation.

\section{Acknowledgments}
		Part of this work is supported by Peng Cheng Cloud Brain II  NPU Cluster in Peng Cheng Laboratory.  We 
		thank our colleagues who provided insight and expertise that greatly assisted the research. We 
		would also like to express our gratitude to the colleagues for their comments that greatly improved the 
		manuscript. 
\small
\bibliographystyle{unsrtnat}
\bibliography{paddle_dist}

\end{document}